\documentclass[11pt,a4paper]{article}
\usepackage[utf8]{inputenc}
\usepackage[square,sort,comma,numbers]{natbib}
\usepackage{graphicx} 
\usepackage{authblk}
\usepackage[many]{tcolorbox}
\usepackage{lipsum}
\usepackage{geometry}
\usepackage[export]{adjustbox}
\usepackage{amsmath,amssymb,subfigure,amsfonts}
\usepackage{epsfig}
\begin{document}
\title{{THE TIMELINE OF GRAVITY 

Let's understand the gravity of this timeline!}}
\author[1]{Arshia Anjum}
\author[2]{Sriman Srisa Saran Mishra}
\affil[1,2]{National Institute Of Science Education And Research Bhubaneswar, Odisha, India - 752050 \and
Homi Bhaba National Institute, Anushakti nagar, Mumbai - 400093}
\date{}
\maketitle
\vspace{-1cm}
\begin{abstract}
    Gravity plays an important part in the experiments and discoveries of the modern world. But how was it discovered? Surely Newton and Einstein were not the only people to observe it and account for it. It had been a long path before the full theory for Gravitation could be formulated with open ends for more add-ons and modifications. All the contributions from across the world and different eras helped in the discovery of gravity as a whole new concept and area of research with a major contribution from the Greeks. This 3 article series lists out the important curves in the carefully carved path of gravitational discovery. The first article summarises the development of interest in the cosmos and the growth of scientific knowledge through ancient theories and observations.
\end{abstract}
\vspace{0.5cm}
{\Large{\textbf{{1. The Ancient Knowledge}}}}
\section*{\large{\textbf{Introduction}}}
\noindent It has often been accepted that science in general, was created by the Greeks. The practicality, intelligence and wholeness of the Greeks, is something which led to the development of logical reasoning and observation. The Greeks had started describing the natural phenomenons and timely occurrence through mythical terms in their literature with the help of Gods and Goddesses. Well, this could be said true for almost all the early civilizations of the Earth, which combined the observation of natural phenomenon and the human belief in God to create a mesmerizing story of what occurred and engrave it in the minds of the then population.\\

But soon with the influence of foreign reasoning, the Greeks started to conquer the expanse of knowledge by giving more importance to mathematical calculations and logical reasoning. The Egyptian mathematics and Chaldean astronomy were studied by Thales, which marked the beginning of their scientific thinking and explanation of phenomena in non-mythical terms . The theories made by Thales about Earth and its suspension in water were taken further by his immediate successor in science,  Anaximander. According to him, the Earth was suspended freely in space and was equidistant from the periphery of the vortex. He was the one taking initiative to introduce the world outside Earth. 

\section*{\large{\textbf{Aristotle and His Views}}}
\noindent Later on carried by Aristotle, the best pupil of Plato, the physical sciences started booming to the next level. He introduced the concept of gravity without actually mentioning it. The Aristotelian way of explaining gravity was that all bodies move towards their natural place when left under no force. For the elements earth and water, that place is the centre of the (geocentric) universe, the natural place of water is a concentric shell around the earth because earth is heavier; it sinks in water.\\

\begin{figure}[ht]
\centering
\includegraphics[width=8.5cm, height=8.5cm]{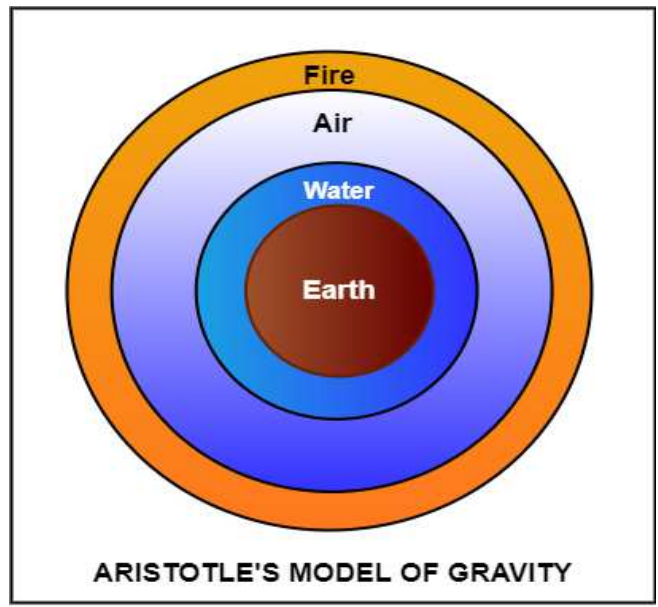}
\caption{Aristotle’s Model of Gravity on Earth depicted by Concentric Spherical Shells.}\label{Aristotle's Model}
\end{figure}

Another point of the Aristotelian physics is that when 2 objects are dropped in the same medium then, the heavier object drops faster than the lighter object, i.e. velocity of the heavier object is greater than the velocity of the lighter object. While the first point can now be regarded as false, we cannot refute the second point in the name of Newtonian physics, because the overall result of the forces acting on the object, like gravity, buoyancy, resistance, etc. leads to this result stated by him. It depends on us if we want to regard him smart enough to combine the result of all the forces and give us the ultimate result, or regard him as not so observant for not being able to resolve all the constituent forces acting on the object and ignore the minimal resistances to state universal laws.

\section*{\large{\textbf{Vitruvius and Density}}}
\noindent Along with the Greeks, Romans are known to have laid the foundation of many breathtaking discoveries in the field of science and technology. During the period of Julius Caesar, one such roman mind namely Vitruvius gave both theory and experimental procedure about many architectural things. He in his book "De Architectura" , described many technologies including the construction of ballista, the interior and exterior design of buildings and monuments etc. He in "De Architectura" postulated that "gravity of a substance depends not on the amount of its weight but on its nature".

\begin{figure}[ht]
    \centering
    \includegraphics[width=8.5cm, height=5cm]{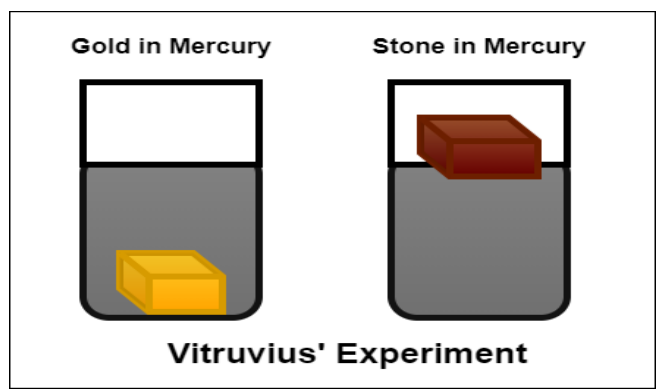}
    \caption{Vitruvius's Observations of the Floating and Sinking of Objects in Mercury.}
    \label{fig:Vitruvius-Model}
\end{figure}

It was based on the observation that when a hundred pound stone is placed on the surface of the mercury, it floats, but when a tiny piece of gold is placed, it sinks down to the bottom of the vessel. From this he concluded that the weight of the object does not influence its pull by gravity, rather its density does. This in today’s world could be related to the Archimedes’ physical law of buoyancy where a body placed on the surface of any fluid either floats or sinks depending  on the difference in the densities of the body and the fluid. This information, though a little misjudged, was instrumental to take the next step in the path to unleash gravity.

\section*{\large{\textbf{Brahmagupta and the Heliocentric Vision}}}
\noindent Indians also gave a huge contribution to the research in the field of astronomy. The most happening period of Indian astronomy started in the 476 CE when Aryabhatta first emerged as an eminent astronomer. In the 7th century, Brahmagupta following the heliocentric vision as defined by Aryabhatta made observations and described this attractive gravitational force by using the term "Gurutvakarshnam" for it. Brahmagupta said that the Earth is the same from all sides and  it is the natural law of the earth to attract all the things towards it. He further said that the earth is the only low thing and said that anything thrown away from the earth in any direction returns back to its surface and never moves upward from the earth. In the language of modern science, we can say that Brahmagupta was aware of the attractive gravitational force but we can also refute his last observation that no object can escape the Earth’s gravitational pull as modern-day rockets and satellite launching vehicles obtain sufficient velocity to overcome Earth’s gravitational pull.

\section*{\large{\textbf{The Medieval Islamic Period}}}
\noindent One of the golden eras of scientific history started in the 11th century. During this period many scientists worked to give their ideas on the various fields of philosophy and sciences. Ibn Sina, gave his famous theory of projectile motion where he explained that the moved object acquires an inclination(mayl) from the mover. Ibn Sina observed the importance of persistent external forces to dissipate the inclination. He also said that the inclination was gained as the body moved against their natural motion. Another researcher, Al Beruni proposed that celestial bodies have mass and gravity just as Earth does, and criticised all the earlier theories for declining this. Abu’ I Barakat Al Baghdadi proposed an explanation of the acceleration of falling bodies, by describing the accumulation of successive increments of power with successive increments of velocity. Ibn Bajjah proposed that every force has an opposite reaction. This could be considered as the early version of the third law of motion described later by Newton. All these discoveries were prominent for the then readers and researchers making way for new concepts and laws.

\section*{\large{\textbf{Buridan and Albert of Saxony)}}}
\noindent After two distinct  11th and 12th century theories on projectile motion given by Ibn Sina and Abu’ l Barakat respectively, there came the marked french philosopher Jean Buridan who postulated a common theory on projectile motion. Here he mentioned that when a person throws something, he gives it a certain impetus which enables that body to move in the direction where it is thrown. It is due to this given impetus that the body keeps on moving after losing contact with the person throwing it and this impetus is proportional to the initial velocity. Like Ibn Sina, Buridan also said that the air resistance and gravity opposes this motion. Finally, gravity dominates and the thrown object moves towards its natural place. The ideas of Buridan were followed by his pupil Albert of Saxony who gave a three stage theory of impetus where he mentioned that initially the thrown object moves in a straight line as the effect of gravity is recessive, in mid way the path deviates downwards as gravity and air resistance opposes the motion. Finally, it moves completely downwards due to the total dominance of gravity.

\begin{figure}[ht]
\centering
\includegraphics[width=\textwidth, height=7cm]{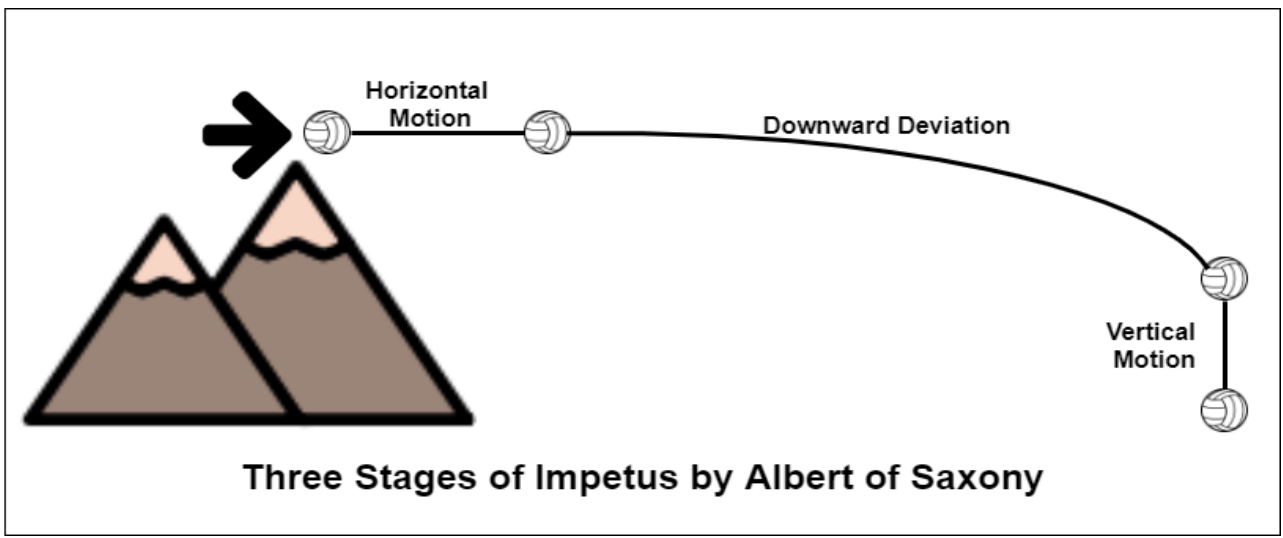}
\caption{Theory of Impetus as given by Albert of Saxony based on the findings of Buridan}\label{Albert of Saxony}
\end{figure}

This theory of impetus is what we now know as momentum. Moreover Buridan also gave a mathematical value to this impetus(momentum) by stating that it is the product of weight(of object) and velocity(applied by the thrower)

\section*{\large{\textbf{Contributions of The Merton College}}}
\noindent The Merton School worked on the kinematic views of motion and found two basic theorems governing the motion of an accelerated object.  First was the mean speed theorem where they said that a body moving with a uniformly accelerated motion covers the same distance in a given time as a body moving for an equivalent duration with a consistent speed same as the mean (or average) speed. 

\begin{figure}[ht]
\centering
\includegraphics[width=12cm, height=9.3cm]{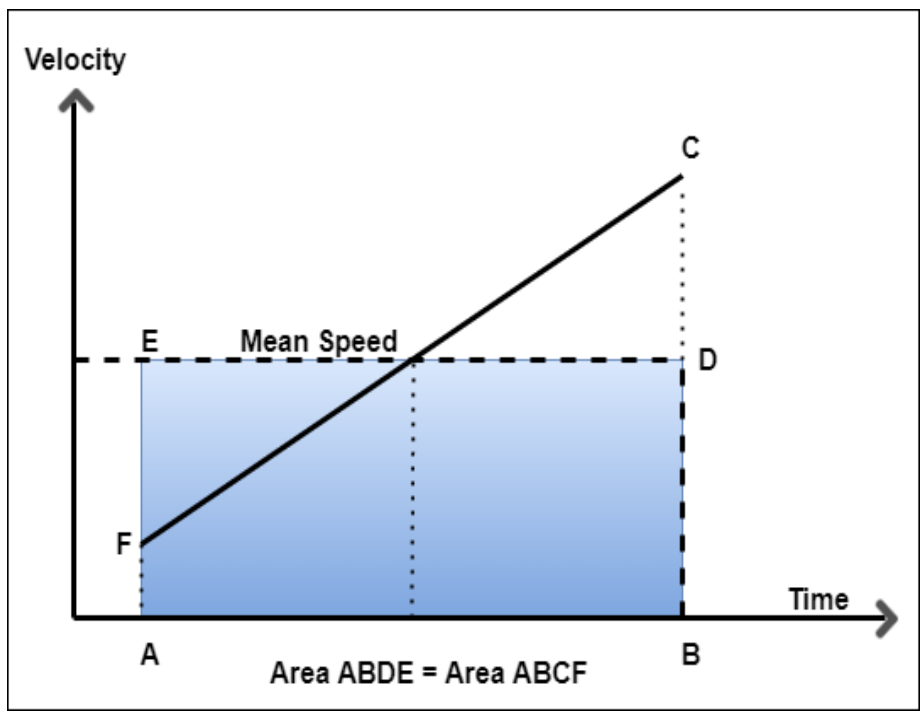}
\caption{The Mean Speed Theorem as Explained by the Merton College}\label{Merton College}
\end{figure}

The second theorem states that the distance covered in the first half of a uniformly accelerated motion starting from rest is the same as a third of the distance covered in the second half. Though these theorems are theoretically correct, they do not completely describe gravity, its source, or its overall unified effects. These theorems have definitely been a very important part of the path to success, thus lessening the gap between our understanding and gravity, But the search is not over yet. Maybe it has just started!

\vspace{0.8cm}
\maketitle

{\Large{\textbf{{ 2. The Beginning of Scientific Experiments}}}}
\section*{\large{\textbf{Introduction}}}
\noindent All the physical concepts have evolved by going through various stages of development consisting of both experimental and theoretical advancements. Both these components go hand in hand. On encountering a problem, we first make observations experimentally and then formulate the plausible theories behind them. It is very important to keep in mind all the aspects while monitoring an experiment to come up with a near perfect theory.\\

Now, an age had started where the experimental proofs were given importance for placing belief in a possible theory. Hence, scientists started coming up with new experimental procedures and ways to record the readings. It is observed that the formation of the concepts in gravity are closely related with the progress in recognising the structure of the universe. Since the Greeks were advanced in Astronomy, even this field did not remain barren from their contributions. \\

Thales followed by Anaximander gave a geocentric view of the solar system where he assumed Earth to be a disk-like structure. Thales said that the earth floats on an infinite ocean. This idea was altered by Anaximander, who claimed the earth to be at the center of an infinite space and surrounded by the rings of fire of sun, moon and stars. This is considered as one of the most revolutionary ideas, because for the first time the Earth was considered to be floating on nothing and without any hard support for itself, thus opening the gate to a lot of mind numbing problems.

\section*{\large{\textbf{Philolaus’ Views}}}
\noindent After various radical modifications and changes came the non geocentric model of the Solar System, as Philolaus, one prominent Pythagorean, said that the universe is made up of the "limiters"(the shape and form) and "unlimited"(air, water and fire) which combine harmonically. According to him the earth and other heavenly bodies revolve around a central invisible fire. His structure also includes a counter-Earth which orbits in such a way that it blocks the view of central fire from the Earth.

\begin{figure}[ht]
\centering
\includegraphics[width=7.5cm, height=7.5cm]{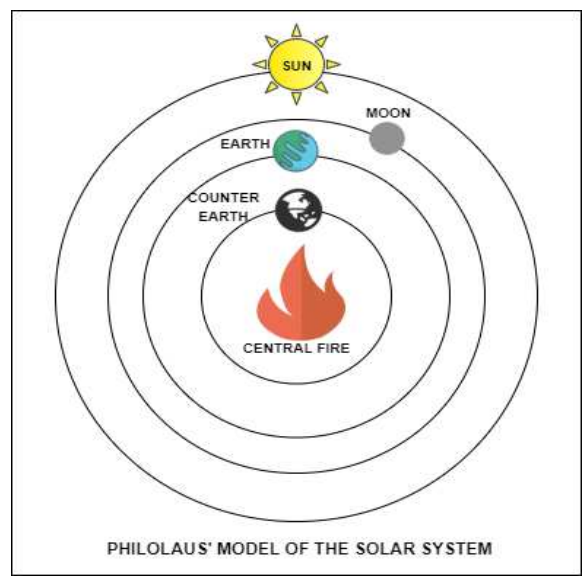}
\caption{This model of our Solar System is given by Philolaus. It assumes a Central Fire to be the pivot which is hidden from the Earth due to the presence of Counter Earth}\label{Philolaus' Model}
\end{figure}

\section*{\large{\textbf{Plato’s Idea and the Aristotelian View of Solar System}}}
\noindent For the first time Plato described the geocentric model of the solar system with Earth at its center. This theory was followed for years, first by Aristotle and then many more. Aristotle was the first to claim that earth is a sphere through various observations. By using the mystical explanations about heaven he succeeded in maintaining the dominance of the geocentric model of the solar system. Years later, another Greek philosopher named Aristarchus came up with the heliocentric model of the solar system where he claimed that the sun was fixed at the center and Earth along with other planets, revolves around the sun and the universe is far more vast than we assume it to be.

\section*{\large{\textbf{Ptolemy’s Model of Solar System (100-200AD)}}}
\noindent Several clashes occurred between the geocentric and heliocentric models of the solar system for centuries. In the second century AD, Ptolemy's geocentric model of the solar system was accepted by the catholic church and was made a doctrine. Here Ptolemy drew a clear picture of the solar system with earth at the center and Moon, Mercury, Venus, Sun, Mars, Jupiter, Saturn and certain other fixed stars, revolving around the earth in ordered perfectly circular orbits.\\

\begin{figure}[ht]
\centering
\includegraphics[width=9cm, height=8cm]{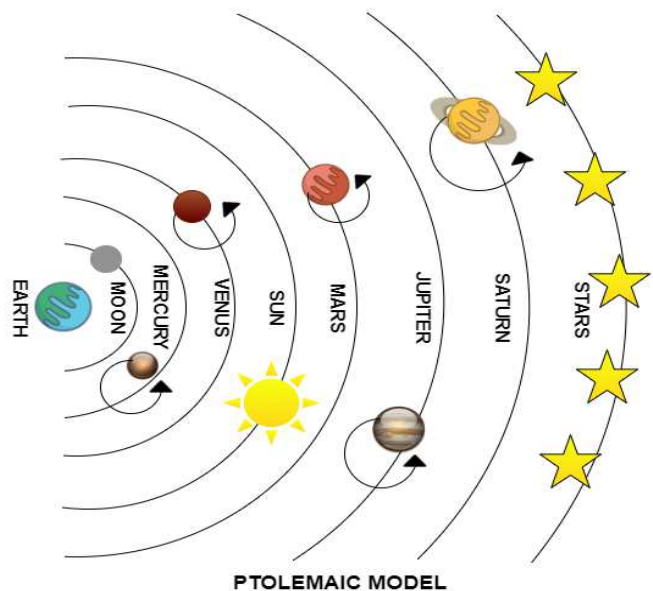}%&
\caption{The Ptolemaic model shows the position of Earth and the other heavenly bodies around the Earth, some revolving in their fixed circular paths and others epicycling in them.}\label{Plotemiac model}
\end{figure}

This model was later challenged by Nicolaus Copernicus who claimed that the Sun is fixed at the center of the solar system and Earth along with other planets, revolves around it. Although this model by Copernicus was suppressed by the catholic churches, it played a great role in driving our views from geocentrism to heliocentrism. 

\section*{\large{\textbf{Copernicus’ Model and Assumptions}}}
\noindent Copernicus did a lot of research on the structure of the solar system in order to discard the flaws that Ptolemy’s model had. He strictly followed heliocentrism and used it as the bedrock for his future works. He, in his notable piece of work "Commentariolus", summarized his heliocentric theory and wrote about the assumptions that he made. One of the assumptions was that the core of the Earth is not the pivot of the universe, but only the pith towards which heavy bodies move and the center of the Earth’s lunar sphere. He was aware of the fact that a force attracts the heavy bodies towards the center of the earth. He also assumed that the motion of the sun and fixed stars as observed by us are not due to their movement but due to the movement of the Earth itself.

\section*{\large{\textbf{Kepler’s and Galileo’s Model of Solar System (1600AD)}}}
\noindent Following the ideas of Copernicus, two renowned astronomers brought a subversive change in the structure of the solar system and succeeded in discarding the ancient geocentric view. In the 17th century Galileo started observing the night sky by using his self-developed telescope and he observed that there are other celestial bodies similar to the moon, that are revolving around Jupiter. He also carefully observed Venus (the brightest of all planets) and saw that it is orbiting the sun, not the earth. On the basis of these observations he claimed that the sun was at the center of the solar system and Earth and other planets orbit the Sun. His theory was silenced at that time due to the excessive power held by the catholic churches.\\

At the same time, Kepler, while observing Mars' orbit in Tycho Brahe’s lab realised that the planets orbit the sun in an elliptical path unlike the perfect circular paths in the Copernicus’ model. He formulated three laws in order to describe the planetary motion around the sun which later on proved to be very helpful in laying the foundations of gravity.

\section*{\large{\textbf{Galileo About Gravity}}}
\noindent Apart from the discoveries related to the cosmos, Galileo had also contributed to the physics behind the vertical and horizontal motion of an object. First by assuming and then by experimental observations, Galileo had  postulated many important theories related to gravity and motion in his unpublished text "De Motu". He was the first to discard Aristotle's theory that the time period of free fall of an object depends on its weight, meaning, heavier objects fall faster than lighter objects. Galileo said that if the air resistance and the buoyant force on a falling object is neglected then all the objects fall with the same acceleration (giving a hint about the constant acceleration due to gravity ‘g’). According to him when two objects of different masses fall from the same height, they take the same time to reach the ground. As written by Vincenzo Viviani, a disciple of Galileo, Galileo proved his statement by dropping two spheres of same masses from The Leaning Tower of Pisa. Galileo also cleared the confusion on the theory of projectile motion. He was the first to say that the projectile motion is a combination of two independent motions-horizontal and vertical- and each of these motions can be studied and observed independently. Although he couldn’t formulate anything about the velocity and time period of vertical motion of an object, he paved the route for many future discoveries related to the vertical and horizontal motion of an object.

\section*{\large{\textbf{Kepler About Gravity}}}
\noindent The pillar for the development of physics behind the celestial motions and gravity was set up by an eminent german astronomer, Johannes Kepler. He was primarily influenced by the heliocentric model of the solar system proposed by Copernicus and gave proper mathematical derivations for many celestial phenomena. While working in Brahe’s lab he was assigned to solve the problem related to the retrograde motion of mars whose reason was not known at that time. He, after thoroughly observing the motion of mars found that unlike the idea of perfectly circular orbits, the planets were moving in elliptical orbits. The shape of the ellipse was known at that time. He also found that the Sun always remains in one of the foci of this elliptical path. Based on this observation, he gave his first law of planetary motion. It stated that  all planets move about the Sun in elliptical orbits, having the Sun placed at one of the foci.\\ 
While working on the problem related to the cause of different velocities of a planet during different phases, he found that the planet moves faster when close to the sun and moves slower when it is far away from the sun. He used this observation to derive the second law of planetary motion which states that an imaginary radial line joining the planet and the sun always traces equal area in equal time period.\\
\begin{figure}[ht]
\centering
\includegraphics[width=11cm, height=8cm]{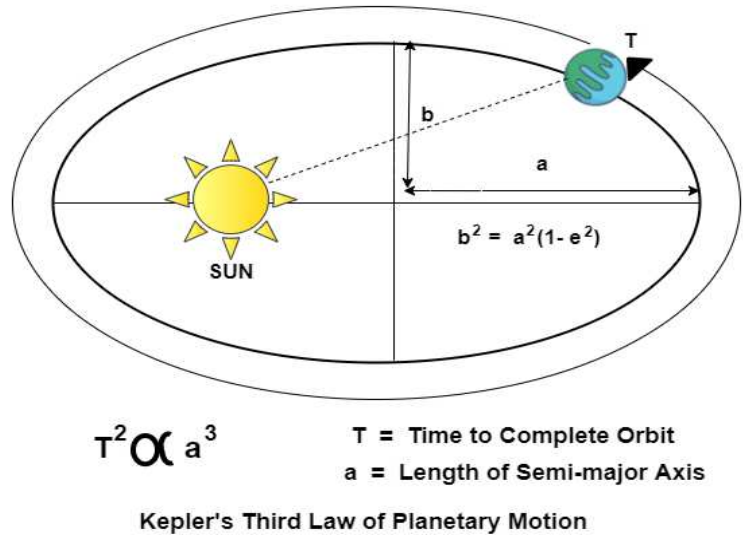}%&
\caption{The third law of Kepler's Planetary Motion shows the relation between the length of semi-major axis of the elliptical orbit of the planet and its time period of one complete revolution around the sun.}\label{Kepler's 3rd Law}
\end{figure}\\

Decades after deriving these two laws, he encountered another fact that as the distance of the planet from the sun increases, the time period of revolution of the planet around the sun also increases. Using this observation he derived his 3rd law which defines the relation between the  time period of revolution of a planet and the radius of its elliptical orbit. This states that the square of the time period of revolution($T$) is directly proportional to the cube of the length of semi-major axis of the elliptical orbit($a$).\\
Mathematically:
$$(T)^2\propto(a)^3$$
Apart from these three planetary laws, Kepler also had some other notable observations as he was the first to identify that the gravitational force is a two body force. Both bodies experience the attraction/pull due to the gravitational force. In a letter to one of his friends he wrote that if we place a stone behind the earth and we assume that the earth and the stone are not involved in any other motions then the stone will move towards the earth and similarly the earth will move towards the stone. These observations and theories were later used by Newton for his research and as stated by NASA in its Kepler’s biography "It was this law, not an apple, that led Newton to his law of gravitation. Kepler can truly be called the founder of celestial mechanics."

\section*{\large{\textbf{Riccioli and Grimaldi}}}
\noindent The 17th century is regarded as an age of great discoveries in science, with innumerable revolutionary ideas and observations. Owing to these findings, Giovanni Battista Riccioli sought to invent a pendulum with an accurate time period of 1 sec, ended up getting close to $0.69\%$ of the desired value. This breakthrough was a major step in the verification of Galileo’s "odd number" rule. Francesco Maria Grimaldi and Riccioli together researched the free fall of objects releasing weights from the Asinelli Tower, using a pendulum to keep track of time and confirmed the dependence of distance of fall on the square of the time taken by the object to reach the ground. The Jesuit Priests themselves did not calculate the value of the acceleration due to Gravity, as we call it today, but when the Newtonian laws of motion were applied to them, it resulted out to be around $9.36 \pm 0.22$  $m/s^2$ ($29.8 \pm 0.7$ $Rmft/s^2$). Though, they could still not resolve the disparity of the dependence of the acceleration on the weight of the object.\\

\begin{figure}[ht]
\centering
\includegraphics[width=14cm, height=8.5cm]{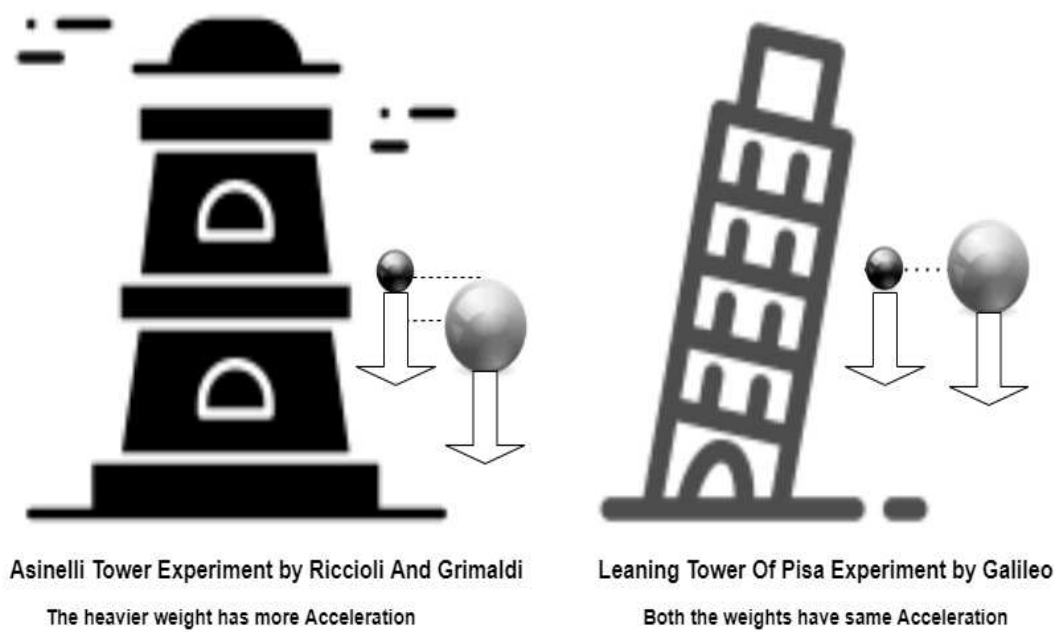}%&
\caption{The image shows the discrepancies recorded in the observation of Riccioli and Grimaldi when compared to that of Galileo.}\label{Riccioli's Experiment}
\end{figure}

As understood by them, if two objects are released simultaneously from a height, the heavier one is expected to descend faster than the lighter one. This statement proved Galileo wrong on the concept, but they didn’t know there were more proofs and disproofs to come. Overall, the concepts of gravity grew to new limits and the experimental procedures grew to new accuracy.

\section*{\large{\textbf{Bullialdus and Inverse Square Law}}}
\noindent Another major step in the development of the basics of gravity was the introduction of the inverse square law. This proposal relates one dependent physical quantity inversely to the square of another independent quantity. It was introduced as a law in the field of Gravity by Ismael Bullialdus in the year 1645. Bullialdus, was majorly working to explain the planetary dynamics and the paths of the interstellar bodies around the Sun. He was skeptical to approve the 2nd and 3rd law of planetary motion given by Kepler and hence kept a view that the Sun exerted an attractive force at the aphelion and a repulsive force at the perihelion, and helped the planets move in a simple, equal and uniform circular motion around the central object, the Sun. Though he could not get the forces right, he is the first person to be credited for the unification of this law in the course to discover The law of Gravity.

\section*{\large{\textbf{Robert Hooke and Borelli}}}
\noindent Robert Hooke gave a lecture on Gravity at The Royal Society in 1666 and another lecture at Gresham College in 1670, in which he clearly gave two principles to ponder upon. The first was that all the heavenly bodies are not only pulled by the central object in lieu of Gravity, but they also exert this gravitational force on each other. The second was that all the bodies which have characteristic simple motion do not tend to curve their path until a continuous external force is applied on them. Both these principles, though really helpful to further the concepts, did not completely prove the universality of the physical quantity. 
Also, Hooke in no way demonstrated the experimental or Mathematical Proof to these principles. He certainly stated that he only hints to these points.

At approximately the same time, Giovanni Alfonso Borelli was also working to explain how the planets moved about their central point. He compared the movement of planets around the sun and their path to the circular motion and orbit of a stone attached to a central axis by a string. Therefore, he stated that the trajectory adopted by the planets was based on the effect of three forces in action.
\begin{figure}
    \centering
    \includegraphics[width=\textwidth, height=7cm]{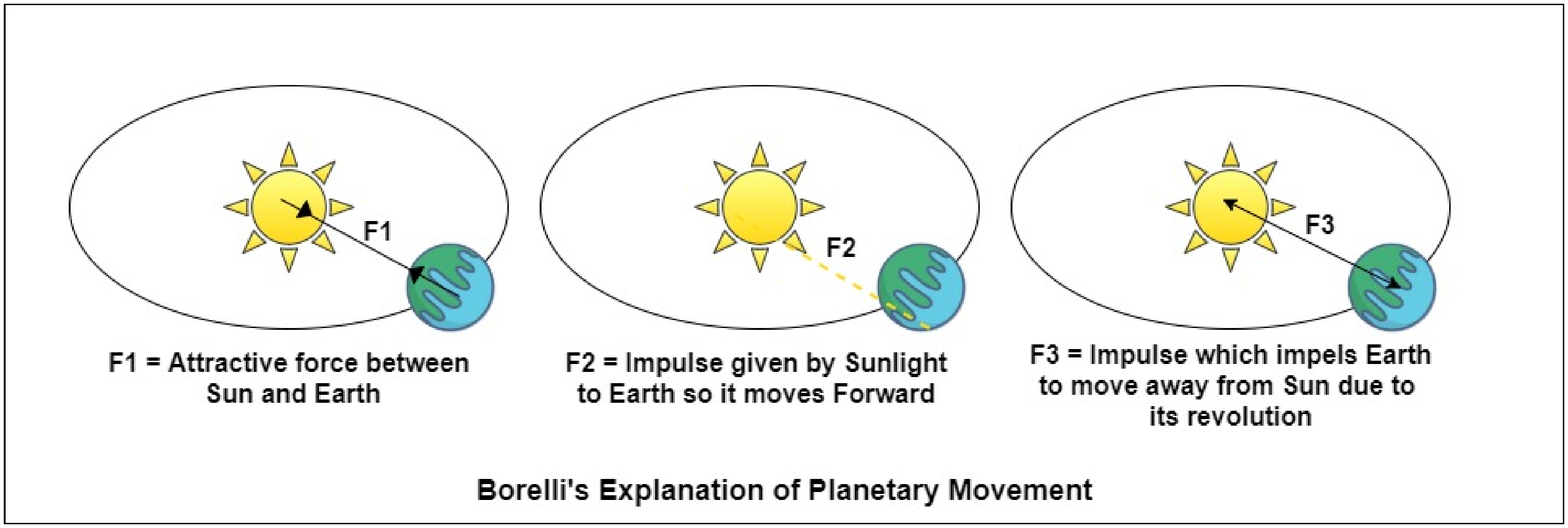}%&
    \caption{The Forces on which the trajectory of a Planet depends as described by Borelli.}
    \label{RoberthookeandBorelli}
\end{figure}
The first force was an attractive force due to which the planet was bound to get closer to the Sun. This he stated to be the "natural appetite of the planets" to move near the Sun. The second force was compelling the planets to move sideways in their orbits due to the corporeal impulse from the Sun. We can imagine the Sunlight exerting a force on the planets as the Sun rotated and thus pushing the planets forward in their orbits. And the third force was also considered to be an impulse which impelled the planet to move away from the Sun due to its revolution around it. The primary result we observe in the research of both these scientists is that they consider gravity to be an attractive force opposing the earlier concept of regarding it as both attractive and repulsive. This proved to be a major breakthrough in the journey to define the universal law of Gravity.

\section*{\large{\textbf{Newton’s Universal Law of Gravitation}}}
\noindent Now the time had come, for one of the Eureka moments to arrive in the field of Gravity. With the inverse square law hinted at and centuries of experiments in free fall, Newton played a pivotal role in joining all the dots and carving out the Universal law of Gravitation. With continuous experiments and observations in classical mechanics, Newton was able to derive the three laws of motion crucial to calculate the acceleration, velocity and position of objects. This was a remarkable feat which helped to acknowledge the fact that an external force is necessary to cause any change in the rest/motion of an object. On the advent of these laws, study of the motion of different celestial bodies started taking a new turn. Newton observed the revolution of the Moon around the Earth, and concluded that there should exist an attractive force between the two celestial bodies. If this was not the case, then the constant force required for a body to overcome its inertia and move in a circular path would not be exerted on the Moon, and the trajectory of the Moon would have been a simple straight line. \\

\begin{figure}[ht]
\centering
\includegraphics[width=8cm, height=8cm]{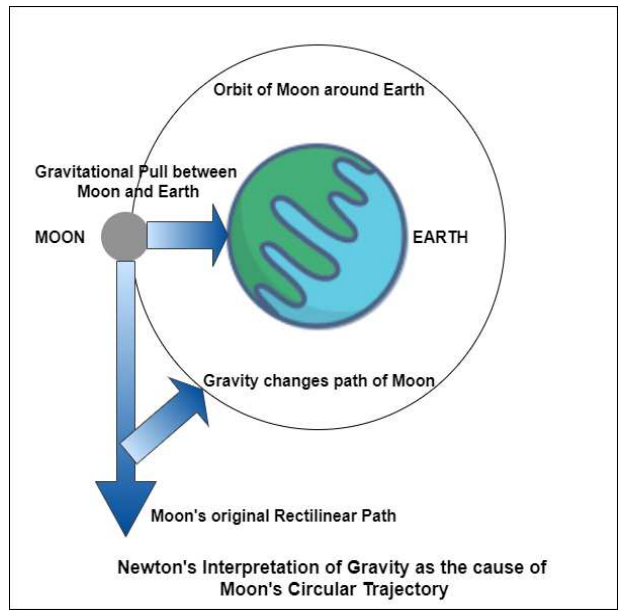}%&
\caption{Newton interpreted the cause of circular trajectory of Moon to be the Gravitational pull of the Earth.}\label{newton}
\end{figure}

Newton also compared the cause of attraction between the celestial bodies to be the same as the cause of an object falling on Earth. In both the cases, Earth exerts a pull on the object and draws it towards itself. But the acceleration with which they come towards Earth is way different. With the inverse square law notioned, Newton was quick to observe that the acceleration of moon towards Earth was approximately $\frac{1}{3600}$ times the acceleration of objects here at the surface of Earth i.e. $\frac{1}{60^2}$, possibly because the distance of Moon from Earth was around $60$ times the radius of Earth. Newton also kept in mind Galileo’s observation that all objects fall towards the Earth at the same acceleration at the surface of the Earth. Hence, he assumed that the force of gravity should be dependent on the mass of the object pulling as well as the mass of the object being pulled, so that when the second law of motion is applied on the pulled object the force by mass ratio becomes constant for all objects on the surface of the Earth.\\
So the 2 major points were :\\
%\justify
$Force~of~gravity \propto product~of~the~two~masses~being~considered$\\
$Force~of~gravity \propto \frac{1}{distance^2}$\\

He assumed a Universal Gravitational Constant $G$ and applied it in the proportionality to get the equation:
$$F = \frac{GM_1M_2}{R^2}$$
%\justify
Where, $G$ is the Universal Gravitational Constant.\\
$F$ is the Force of Gravity.\\ 
$M_1$ the mass of the first object.\\
$M_2$ the mass of the second object. \\
$R$ the distance between the centres of the two objects.\\

But then arose a problem: How could we use this equation for an object which cannot be considered a point mass and has dimensions in it. It was resolved by using vectors. Every particle of one object exerted a force of gravity on every particle of the other object. And hence the net force of gravity was the vector sum of all these small forces. This law, with the help of Newton's third law of motion, proved that the force exerted by Earth for pulling an object was equal to the force exerted by the object to pull the Earth towards itself. Still we do not see the Earth moving to a dropped object. As the Mass of Earth is huge, the force by mass ratio comes out really small and hence the acceleration of the Earth is really meager. While the force by mass ratio of the object is large enough to be observed and hence we see the object move towards the Earth.\\ 
Then came the question of gravity in celestial bodies. Newton explained that this law seems approximately applicable in the heavenly bodies as the distance between two celestial objects is far greater than the size of the bodies themselves. Hence, with respect to the distance between them, the bodies are considered to be point objects with all their mass concentrated at the centre of mass. This force of gravity between the celestial bodies helps them to move in specific circular orbits around each other. Thinking of our solar system, every planet is attracted to the Sun by gravity, and every planet exerts a gravitational pull on every other planet in the outer world. This keeps a balance between all the forces and hence helps the celestial bodies to revolve in orbits around the Sun. The Newton’s Universal law of Gravitation was advantageous to explain the trajectories of all the planets in the solar system except mercury which displayed small deflection from the law, but ample to be noticed by astronomers.\\ 
Some of the most remarkable points of Newton’s Universal Law of Gravitation were that this was one of the first theories to include the concept of a centripetal force rather than a centrifugal one. This is a much debated topic on whom this idea belongs to. Some say Newton was suggested this by Robert Hooke in one of his letters to him while he was writing his book Principia Mathematica. Newton claimed to prove Kepler’s planetary laws with the help of his law of Gravitation. So we can say if Kepler was the one to give the idea, Newton was the one who proved them. Also, Galileo’s observations were affirmative of his law. This law had revolutionized the study of Gravity around the world pushing the level of understanding up a step. Though it could not help identify why there seemed to be an anomaly in the trajectory of Mercury, most of the problems were solved using these equations and it caused yet another marvel in science.

\section*{\large{\textbf{Planet Vulcan to the Rescue}}}
\noindent So the inconsistency in the orbital trajectory of the planet Mercury was noticed by all. The planet clearly did not follow the path marked by the astronomers using Newton's Universal law of gravitation and laws of  motion. So, they assumed there needs to be something present in the area which also exerted another gravitational pull on Mercury , causing its observed path to deviate from the theoretical path through perihelion precession by 43 arcseconds. In 1859, the French scientist, Urbain Le Verrier proposed a theory in which there was a hypothetical planet revolving around the Sun, whose orbital radii was shorter than that of Mercury. Since this planet was so close to the Sun, it was named by Verrier as planet Vulcan, wherein Vulcan is the God of Fire. As Verrier was successful in finding the planet Neptune to compensate for the inconsistency in the trajectory of planet Uranus, this theory of his was believed by many and from then began the search for Vulcan. Verrier received many letters confirming the existence of black small planet nine, but he could never in his lifetime see the proof for the existence of this planet. Vulcan, the so-called cause of deviation in Mercury’s trajectory. Or is it? The hunt for Vulcan continued.

\vspace{1cm}
\maketitle

{\Large{\textbf{{3. The Advancements in Gravity}}}}
\section*{\large{\textbf{Introduction}}}
\noindent To account for the deviation in Mercury's trajectory, the search for planet Vulcan continued. On the other hand, the value of Universal Gravitational Constant was yet not derived. Newton had tried to give a theory for finding the value of $G$ on a mountain, but he himself could not accomplish the task of calculating it. Also, some scientists were still skeptical of Newton’s Model of Universal Gravity. During the late 17th century, the theory of Huygens about gravity clashed with that of Newton as their way of approaches were different. Newton said that gravity is the mutual force of attraction between two bodies in the universe. While Huygens used classical mechanics to explain the phenomenon of gravity. He followed the vortex theory of Rene Descartes, which assumed the space to be non-empty and completely filled with circular rotating matter. The circular motion gives rise to vortices in the space. And due to the centrifugal force emerging from these circular motions, the fine matter moves towards the edges of the vortices. While the rough matter resists the centrifugal force resulting in a pressure applied by the fine matter on rough matter, forcing the rough matter to move towards the center of the vortex. Huygens called this inward pressure as gravity. He also modified Descartes' theory and used spherical vortices instead of cylindrical ones to explain gravity. \\

Huygens stated that the celestial objects must have holes through which Aether (the imaginary substance which was earlier regarded as the matter filling the space) can penetrate. According to him the force applied by the fine matter on the rough matter in the direction of the center of the vortex was equal to the centrifugal force that was considered to have moved the fine matter to edges of the vortices. He also derived the mathematical formula for this inward force gravity which today is known as the centripetal force of attraction. This mathematical expression enabled the conversion of Kepler’s third law of planetary motion into the inverse square law of gravitation easily.  Huygens was only able to give an acceptable explanation for terrestrial gravity. His mechanical explanation was unable to explain the elliptical motion of planets and other celestial bodies like comets around the sun as they failed to correctly give the reason behind why the aether was not applying retarding force on the movement of planets.  This was better explained in Newton’s model, and hence it received more acceptance.

\section*{\large{\textbf{Universal Gravitational Constant}}}
\noindent
While the universal law of gravitation had now become the explanation of the mysterious gravity, It was still incomplete without the Universal Gravitational constant ($G$). Though Newton could not determine the value of G, he had predicted the order of the Universal Constant to be around $10^{-11}$ $m^3/kg$ $sec$. Pierre Bouguer and Charles Marie de La Condamine, during their "Peruvian Expedition" tried to weigh mountain Chimborazo with the help of a pendulum. They used the fact that the deviation in the pendulum from vertical position would be directly proportional to the ratio of the density of the mountain to the density of Earth. They implied that as the deviation is very meagre, Earth cannot be a hollow sphere, thus indicating the measurement of Earth’s density. On noting the success of this experiment, the Royal Society decided to conduct the same experiment on mountain  Schiehallion under the surveillance of Charles Hutton and Reuben Burrow who measured the deviation in the pendulum to be around $11.6$ arcseconds and predicted the density of Earth to be approximately $4.5$ times that of water. And since, the Earth’s density indirectly affects the Gravity by the relation ,
$$\frac{\rho_{earth}}{\rho_{mountain}}=\frac{V_{mountain}\cdot{R_{earth}}^2}{V_{earth}\cdot{d}\cdot{\tan}\theta},\hspace{2cm}[1]$$
where,\\
$\rho_{earth}$ = Density of Earth,\\
$\rho_{mountain}$ = Density of Mountain,\\
$V_{mountain}$ = Volume of Mountain,\\
$R_{earth}$ = Radius of Earth,\\
$V_{earth}$ = Volume of Earth,\\
$d$ = Distance between the Centre of Mass of Mountain and the Pendulum,\\
$\theta$ = Angle of deviation,\\

This was the first successful indirect method to calculate the value of universal Gravitational Constant. The direct reading was first taken by Henry Cavendish in his Torsional Balance Experiment. His main aim was to use the torsional balance apparatus devised by his friend John Michell to record the density of the Earth, but ended up getting a direct measurement method for Universal Gravitational Constant In the model, Cavendish measured the distance $d$ (distance between the equilibrium position of the small oscillating mass $m$ and the fixed heavier mass $M$), the deviation angle $\theta$ (deviation between the equilibrium position when heavier mass $M$ is there and when it is not there) and the oscillation frequency $T$ of the oscillating masses.

\begin{figure}[ht]
\centering
\includegraphics[width=5cm, height=7.5cm]{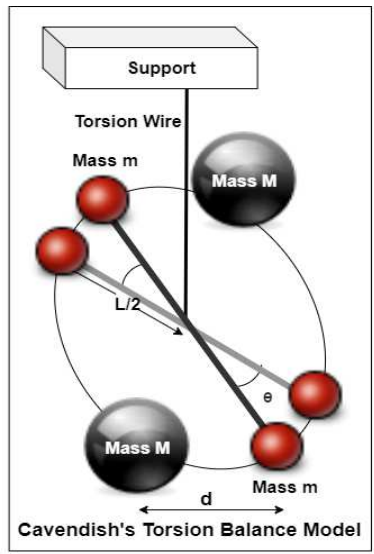}%&
\caption{Cavendish used the Torsional Balance Apparatus made by his friend Michell to directly evaluate the value of Universal Gravitational Constant}\label{Cavendish's Experiment}
\end{figure}

The derived equation for $G$ is:\\
$$G=\frac{2{\pi}^2L\theta{d}^2}{T^2M},\hspace{2cm}[2]$$
where,\\
$G$ is the Universal Gravitation Constant,\\
$L$ is the length of the torsion bar,\\
$\theta$ is the angle the bar turns,\\
$d$ is the equilibrium point distance between $M$ and $m$,\\
$T$ is the oscillation frequency,\\
$M$ is the mass of the larger object.\\

He was successful in recording a precise value, with just a deviation of $1.6\%$ in the density of the Earth. And hence, the first direct successful measurement of universal Gravitational constant took place. John Henry Poynting revised the readings of Cavendish to a closer value with just $0.2\%$ deviation from the modern value.\\ C. V. boys and Carl Braun calculated the value of \textit{G} to be 
% $6.683 \times 10^{−11} m^3 {Kg}^{-1} {sec}^{-1}$
. The modern value of the gravitational constant has been set by Paul R Heyl as
% $6.670 \times 10^{−11} m^3 {Kg}^{-1} {sec}^{-1}$ 
with a relative uncertainty of $0.1\%$

\section*{\textbf{\Large{The Beginning of Modern Physics}}} 

\large{The problems of Gravity were readily being solved by Newton’s theory and the anomaly in the trajectory of Mercury was still being researched. However, the scientists could still not locate planet Nine hypothesised by Verrier to compensate for the anomaly. So, this posed as a very big drawback of Newton’s theory of gravity. Somewhere around 1905, a genius named Albert Einstein, started one of his revolutionary thought experiments, where he imagined what would happen if he moved at a very high speed, a speed comparable to that of Light. Einstein was deeply influenced by Maxwell’s electromagnetic rules and Newton’s idea of mechanics. But, he somehow found a point of disagreement between the  theories given by these two. Newton’s laws of mechanics hold true for all the inertial (neither accelerating, nor rotating) frames of reference.\\

So, we observe all the physical phenomena in the inertial reference frames, and hence can apply Newton’s mechanical equations to describe their dynamics. Newton believed that the laws of physics do not change based on the velocity of the observer. This meant that, for all the reference frames, either stationary or moving with a constant velocity with respect to a stationary frame, would follow the laws of physics in the same way and provide similar end results and observations. The electromagnetic wave theory given by Maxwell and further transformed by Lorentz stated that fluctuating electric and magnetic fields travel through vacuum at the speed of light. This embraced the fact that the electromagnetic laws applied to all inertial frames of reference.\\

\begin{figure}[ht]
\centering
\includegraphics[width=11cm, height=6cm]{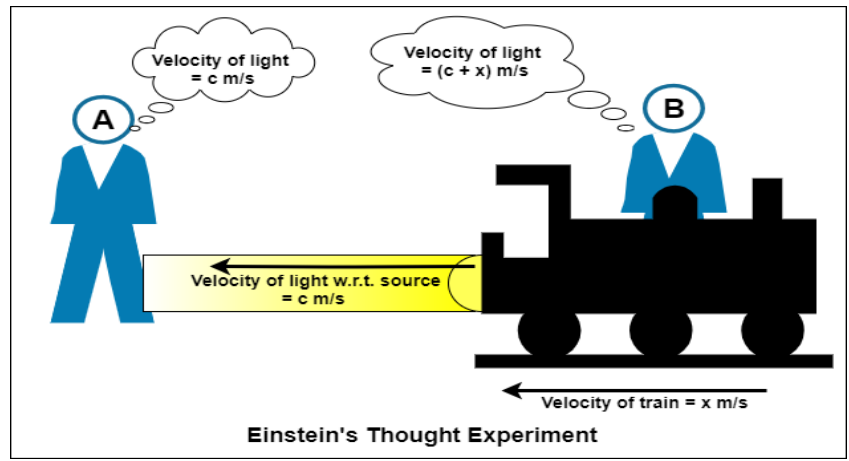}%&
\caption{Einstein's thought experiment of a hypothetical train violating Maxwell's laws of Electromagnetism.}\label{Einstein's thought experiment1}
\end{figure}

Keeping these two laws as the basis, Einstein came up with a thought experiment, where he encountered a disagreement between these laws. He imagined a train in vacuum moving at a constant speed, which is comparable to the speed of light, let it be $x$ m/s.There is a light source on the engine of the train, which emits light in the forward direction. A person standing on the platform, let's say person $A$, would observe the speed of light from the source to be \makebox{c = $3$ x $10^{8}$ m/s}. But a person present in the train, let's say person $B$, will observe the speed of light to be $(c +x)$ m/s. This difference in the velocity of light observed by both $A$ and $B$ shows that Maxwell’s law is violated in this experiment.\\

Einstein concluded that the two frames of reference would be experiencing time differently. He came up with the hypothesis that time is relative to the observer. He gave one more example to suffice this fact.
\begin{figure}[ht]
\centering
\includegraphics[width=14cm, height=5.5cm]{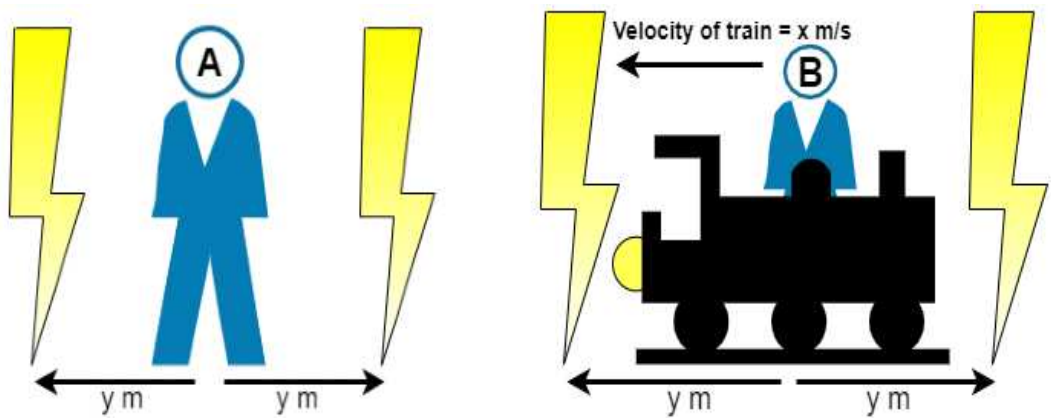}%&
\caption{Einstein's thought experiment of a lightning striking the two equidistant ends of train and platform.}\label{Einstein's thought experiment merged}
\end{figure}
 
Suppose that when the mid point of the train is at the middle of the platform, lightning strikes at two places equidistant from the middle. Person $A$ standing on the middle of the platform would, according to his observations, say that both the lightning struck the ground simultaneously, as light would travel equal distances to reach the observer. But for person $B$, as he is moving forward with a speed comparable to that of light, the lightning striking the front end would be observed first, as it would travel a lesser distance to reach the observer, while the one at the back end would be observed later, as it would have to cover a larger distance. So, a phenomenon recorded by $A$ as simultaneous, will be recorded by $B$ with a small interval of time. With this Einstein concluded that time is relative, and gave his two postulates of special theory of relativity.\\

The first postulate is called the principle of relativity. It states that the laws of Physics remain unchanged in all the inertial frames of reference. The second postulate is called the principle of constancy of velocity of light. It states that the velocity of light in a medium remains consistent for all observers, regardless of the dynamics of the source or observer. These two postulates led the foundation of two completely new topics for the world to ponder upon: Time dilation and length contraction. Time dilation can be said in abstract terms as time moves in a sluggish way when the clock is in relative motion to the observer. Since we now know through special relativity that the speed of light always remains constant for all frames, let us consider one more thought experiment. Consider person $A$ on a movable platform assisted with a clock and two plane mirrors fixed at a definite length $l$ which are facing each other. The axis of mirrors is perpendicular to the direction of motion of the platform. The mirror at the base has a light source attached with it. When the light source is switched on, the beam of light travels straight to the other mirror and gets reflected back to the first mirror and this goes on. Now consider another observer, person $B$ standing on the ground with another clock. When the platform moves at a speed $v$ comparable to that of light, let the time taken by light to reach the opposite mirror as observed by person $A$ be $t_1$. So the distance light travels one way is $ct_1$. From Person $B$’s perspective, due to the movement of the platform, the light emitted by the source does not reach the other mirror by travelling perpendicularly. Rather it traces an oblique path. Let the time taken by light to reach the opposite mirror as observed by person $B$ be $t_2$. So the distance light travels one way is $(vt_2)^2$ + $l^2$. But since it is a physical system, we cannot negate that the distance travelled by light can also be written as $ct_2$. Equating the two possible expressions we get,  
$$t_2= \frac{1}{\sqrt{c^2 - v^2}}\hspace{2cm}[3]$$
$$\frac{1}{c} = t_1\hspace{3.3cm}[4]$$
$$t_2 = \frac{t_1}{\sqrt{1 - \frac{v^2}{c}}}\hspace{2.1cm}[5]$$ 
As $v$ is always less than $c$, the denominator is less than $1$ and hence $t_2$ \textgreater $t_1$. This proves that in inertial frames that move with high speed, the time dilates with respect to the inertial frames at rest. 
\begin{figure}[ht]
\centering
\includegraphics[width=14cm, height=5cm]{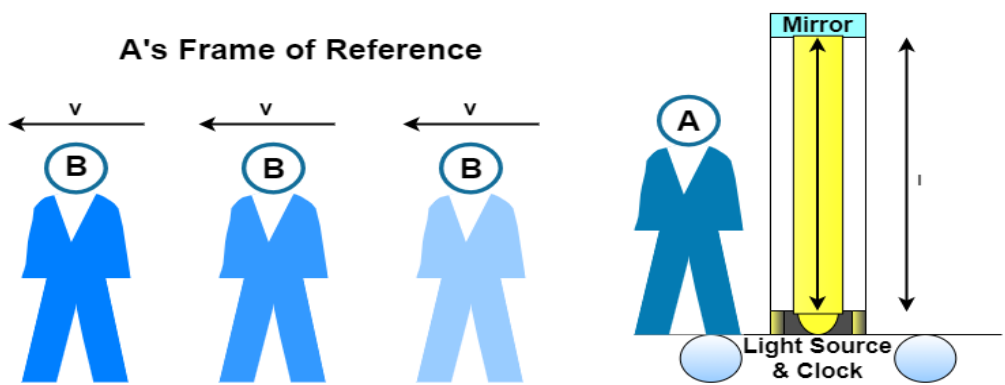}%&
\caption{Person A observes person B moving backwards (shown in fading colours) and the light ray move in straight line to the opposite mirror.}\label{Einstein's thought experiment2_1}
\end{figure}

\begin{figure}[ht]
\centering
\includegraphics[width=14cm, height=5cm]{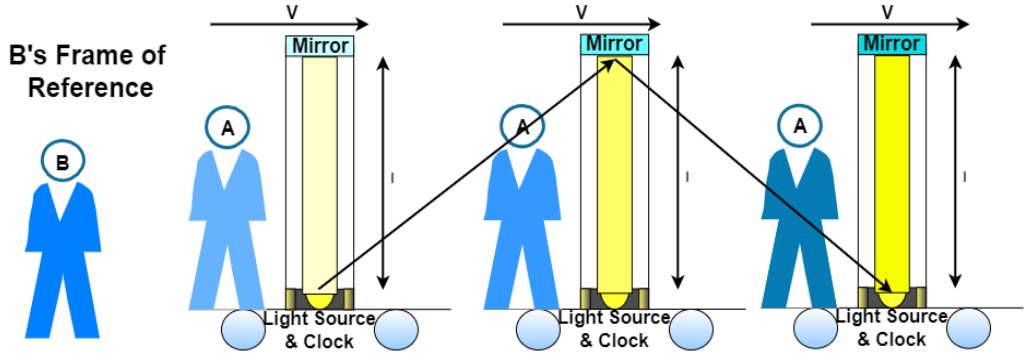}%&
\caption{Person B observes person A moving forward (shown in fading colours) and the light ray move in an oblique trajectory to reach the opposite mirror.}\label{Einstein's thought experiment2_2}
\end{figure}
Since space and time are relative, there is no absolute space or absolute time in the universe. When at high speed the time dilates, it also causes a change in the length of the system. Let person $A$ observe the length of the system to be $l_1$. Then, $l_1$ = $ct_1$. The length observed by person $B$ is $l_2$. Then, $l_2$ = $ct_2$.

Substituting the value of $t_2$ we get:

$$l_2= \frac{ct_1}{\sqrt{1 - \frac{v^2}{c^2}}}\hspace{2cm}[6]$$
$$l_2 = \frac{l_1}{\sqrt{1 - \frac{v^2}{c^2}}}\hspace{2cm}[7]$$
$$l_1 = \frac{l_2}{\sqrt{1 - \frac{v^2}{c^2}}}\hspace{2cm}[8]$$

As $v$ is always less than $c$, the square root factor is less than $1$ and hence $l_1$\textless$l_2$. This proves that in inertial frames that move with high speed, the length of the system contracts with respect to the inertial frames at rest.}

\section*{\textbf{\Large{General Relativity}}} 

\large{Until the early 19th century, people were aware about the physical laws that hold only for the bodies which show uniform rectilinear, non-rotary motion, that is for the inertial reference frame. The special theory of relativity has shown us that when we use these inertial references (generally know-n as Galilean reference bodies) for formulating the general laws of nature , both these references give us the exact same interpretation of the physical law. In the early 19th century Einstein thought of explaining a universal law for gravity by weaving his special theory of relativity and Newton's law of gravitation. His objective was to take into account the effect of gravity on the motion of the objects in the universe. The first thing which Einstein pondered on was the equality between the gravitational and the inertial mass. For this he introspected on one of his thought experiments where he imagined a man in free fall. He then realised that during free fall, a body only experiences the gravitational force by the Earth as it has no physical contact with any other object or with the surface of the earth for other forces to act on it. \\

\begin{figure}[t!]
\centering
\includegraphics[width=10cm, height=7.5cm]{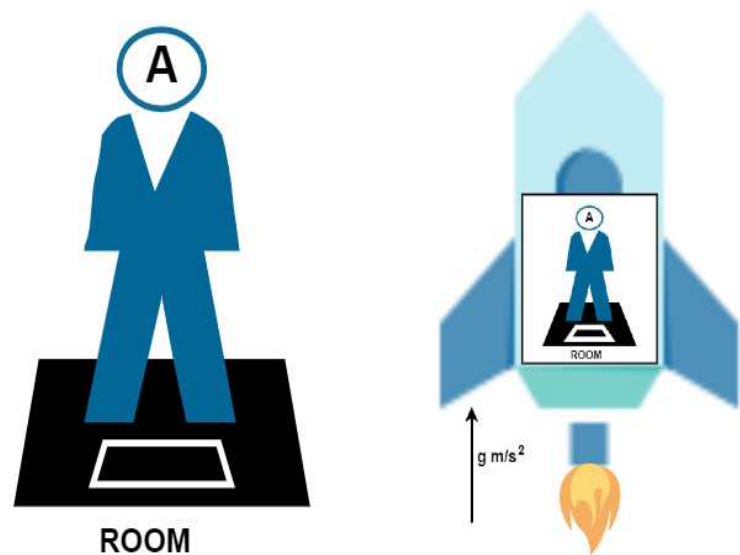}%&
\caption{The figure shows two cases of Einstein's thought Experiment that he used to determine the equality in the Inertial Mass and Gravitational mass.  }\label{Einstein's thought experiment2-3}
\end{figure}

So if we neglect the effect of air resistance we can compare the free falling man with a man in space. At first he imagined a man inside a closed room (without any window) on earth’s surface. In this case the weight of the man will be equal to his actual weight. Then in the second case he imagined the same man to be inside a rocket accelerating upwards with $9.8$ $m/s^2$. In the second case, the weighing machine should give the same measurement as it gave when used on Earth. The reason was that the upward acceleration of the rocket should have the same effect as  gravitational acceleration acting on the man should produce. This notion gave rise to his renowned equivalence principle, which stated that the gravitational and the inertial mass are equal and there is no difference between the motion in an inertial frame and the motion governed by the gravitational force. He then concluded that these two types of motion are equivalent and indistinguishable.\\

Secondly, he thought of one more hypothetical scene which brought some radical changes in physics. He thought of a man inside a rocket moving with an upward acceleration of $9.8$ $m/s^2$. The man points a flashlight from one side of the room to the opposite wall. He realized that the path of the light will bend as the particles of light would be under the effect of net downward acceleration with respect to the room. The height of the beam of light (from the bottom of the room) near the source will be more than its height when it strikes the opposite wall. He then imitated the same scene in a room on the earth’s surface. Here also, he realised that the path of light must appear bent or else it will violate the equivalence principle. But then a question came into his mind that if light always takes the shortest path between two points (Which is said to be the straight line path) then why is the path of light bending in these cases. He finally realized that in the presence of mass and energy, gravity causes curvature of space and the shortest possible path is not a straight line rather a curved line. Defining these curved geometry mathematically was indeed a  challenging task for Einstein. He seeked help from one of his college mates and a very good mathematician Marcel Grossman. Both of them figured out that the curved spacetime can be depicted using Riemannian geometry.\\
\begin{figure}[ht]
\centering
\includegraphics[width=13cm, height=6cm]{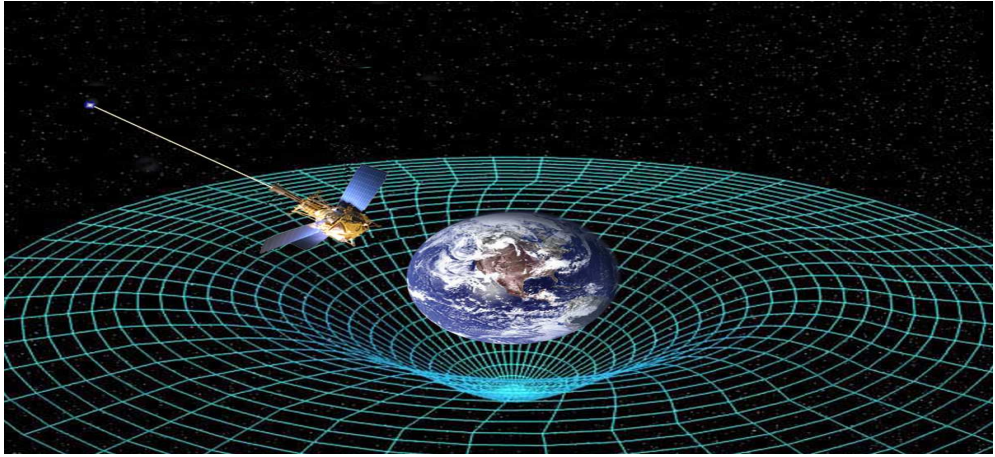}%&
\caption{The figure shows the space-time wrapped around the earth. The presence of massive object (here earth) causes distortion in the space-time and curves it.\\(Image: \copyright \hspace{1mm}NASA)}\label{Einstein's thought experiment2-4}
\end{figure}

Then Einstein thought of a case where he tried to relate time with the curved space in the vicinity of high mass. According to his special theory of relativity, the speed of light is constant in all frames and is equal to $c$ = $3$ x $10^{8}$ $m/s$. When light passes through a curved space caused due to massive celestial objects, it traverses the curvature of the object in order to travel across it. But when light travels in free, undistorted space it travels in a straight line. So for the Special theory of relativity to be true the light must take more time to travel in curved space (as it covers more distance) than the time it takes to travel in free space. Einstein inferred that space and time are wrapped into a fabric of 4- Dimensional spacetime which becomes curved in presence of high mass. This is nothing but the interaction of the mass with spacetime.\\
\\
He formulated these observations and the effect of gravity on space-time curvature mathematically using Riemannian geometry and other mathematical tools and gave a set of partial differential equations called the Einstein’s Field Equations. These equations create a relationship between the space-time curvature and the energy and momentum of matter and radiations present all around the space. 

The equation is :-
$$G_{\mu\nu}+\Lambda{g_{\mu\nu}}=\kappa{T_{\mu\nu}}\hspace{2cm}[9]$$
Here the $G_{\mu\nu}$ is the Einstein tensor and can be defined as the difference between $R_{\mu\nu}$ and $1/2Rg_{\mu\nu}$ that is $$G_{\mu\nu}=R_{\mu\nu}-\frac{1}{2}Rg_{\mu\nu}\hspace{2cm}[10]$$ where $R_{\mu\nu}$ is Ricci curvature tensor, $R$ is Scalar Curvature and $g_{\mu\nu}$ is metric tensor.\\
$\Lambda$ is the cosmological constant.\\
$\kappa$ is Einstein’s gravitational constant which is defined as $$\kappa = \frac{8\pi{G}}{c^4}\hspace{2cm}[11]$$
$T_{\mu\nu}$ is the stress energy tensor.\\
The left expression, that is $G_{\mu\nu}+\Lambda{g_{\mu\nu}}$ represents the space time curvature and the right expression, that is $\kappa{T_{\mu\nu}}$ represents the matter-energy content of space-time.

Einstein’s general theory of relativity helped in explaining various unexplained anomalies such as the Mercury’s orbital precession. After decades of brain cracking research on the oddity of Mercury, finally came a theory inclusive of all such processions as well as the classical concepts of gravity. This proved to be one of the major factors for the success of this theory. The spacetime can’t be seen or measured but many phenomena had been explained using the wrapping of space and time, for example Gravitational lensing, Frame Dragging of space-time around rotating bodies, gravitational redshift, gravitational waves etc. }

\section*{\textbf{\Large{Gravitational Redshift}}} 

\large{Now, let us think about the photons emitted from a massive celestial body far away from the Earth. Then, while the photon comes towards the Earth, it will have to overcome its own gravitational field, as well as cross the gravitational field of other celestial bodies. To resist the gravitational force and move towards the Earth, it needs to expend energy.\\

When a photon releases energy, it leads to the lowering of its frequency. As Einstein’s principle of constancy of velocity states that light has a consistent speed in vacuum, the decrease in frequency will correspond with the increase in the wavelength of the photons, thus shifting them from the blue region to the red region of the spectrum. This phenomenon is known as the Gravitational Redshift.
\vspace{5mm}

\begin{figure}[ht]
\centering
\includegraphics[width=\textwidth, height=4cm]{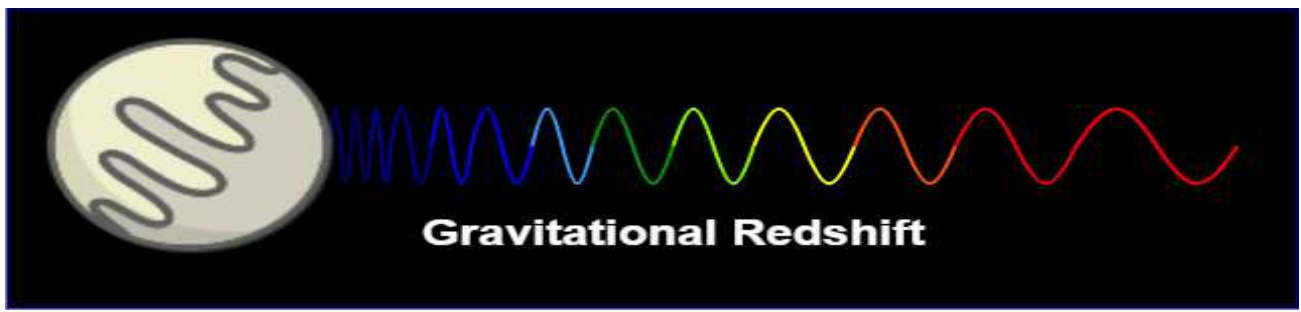}%&
\caption{Gravitational redshift of a photon trying to overcome the gravitational field of a planet}\label{Einstein's thought experiment2-5}
\end{figure}

Exquisite proof of gravitational redshift was given by Robert Pound and Glen Rebka in their 1959 experiment. They placed a source of gamma rays at the base of the top of the tower and a receiver at the bottom of the tower. When gamma rays were ejected from the source, they moved towards the Earth’s centre, i.e., towards a high gravity area, and hence showed a blue shift. Similarly, when the receiver was placed at the top and the source at the bottom of the tower, it showed a red shift, as gamma rays released energy to go away from the high gravitational field. This gave Einstein’s prediction the much required substantial proof of gravity affecting light.}

\section*{\textbf{\Large{Frame Dragging}}} 

\large{According to the general theory of relativity, Einstein proposed that spacetime is curved around massive objects, thus accounting for the gravitational effect. Now as a celestial body rotates, it causes an extra drag in the spacetime curve, thus distorting it to an extent.\\

The circular velocity of the celestial bodies orbiting the spacetime of a particular mass is altered due to the rotation of this mass. This frame dragging has been observed prominently around black holes and stars as they have enough mass concentrated to cause a significant drag due to their own rotation in the path of bodies revolving around them.\\ 

\begin{figure}[ht]
\centering
\includegraphics[width=7cm, height=7cm]{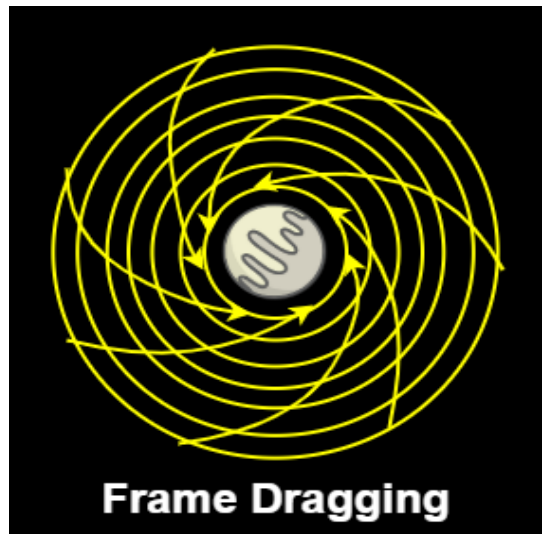}%&
\caption{When a planet rotates about its axis, it cause a drag in the nearby spacetime thus distorting it a little }\label{Einstein's thought experiment2-6}
\end{figure}

The first frame dragging effect was derived by Josef Lense and Hans Thirring in 1918. Hence it is also known as the Lense-Thirring effect. The proof of frame dragging around Earth has been confirmed using the geodesic satellites by receiving laser signals from them in periodic intervals of time in various stations around the Earth and accurately measuring the drag in their path. The Earth’s drag, though feeble due to its weak gravitational field, is still observable, and hence proved Einstein’s prediction in the General theory of Relativity.}

\section*{\textbf{\Large{Discovery of Black Hole}}} 

\large{Black hole is a region in spacetime where even light cannot escape due to the presence of very strong gravity. Einstein predicted the existence of the black hole in 1916 when a famous physicist named Karl Schwarzschild found the first non-trivial solution of Einstein's field equation.\\
These solutions framed the foundation for the discovery of the blackhole. Blackholes are created when compact objects such as super giant stars undergo gravitational collapse. A star remains stable only if the radiation emitted due to the fusion of hydrogen atoms in its core establishes an equivalence with the gravitational force. If the star has more mass, then it allows the fusion of more heavier elements in its core until it reaches iron. But the fusion process that creates iron does not generate any energy and more mass gets accumulated in the core that increases the intensity of the gravity relatively. So the balance between the gravitational force and the radiation breaks and the core collapses. The core implodes and the entire mass of the core collapses into a black hole. The term Black hole was used for the first time by an American astronomer John Wheeler and in 1971 the first ever black hole Cygnus X-1 was observed.}

\section*{\textbf{\Large{Gravitational Lensing}}} 

\large{In this natural phenomenon, when a light ray passes through a very massive celestial body like a blackhole, galaxy cluster, etc.  it gets bent due to the curvature of spacetime that is caused by the heavy mass. This bending of light causes the massive body to act as a lens for the celestial objects placed behind it. This phenomenon was initially experimentally verified by Arthur Eddington and his collaborators in 1919 when they observed a star positioned behind the sun during a solar eclipse. Previously, they were observing the star to be situated behind the Sun. But on the day of the total solar eclipse, they noticed the starlight being deflected by the Sun and the star appeared in an apparent deflected position. This observation worked as an experimental proof of Einstein’s General Theory of Relativity. The first gravitationally lensed object was discovered in the year 1979. It was a Quasar that appears as two images due to the gravitational lensing caused due to galaxy YGKOW G1.\\
\begin{figure}[ht]
\centering
\includegraphics[width=5.5cm, height=6cm]{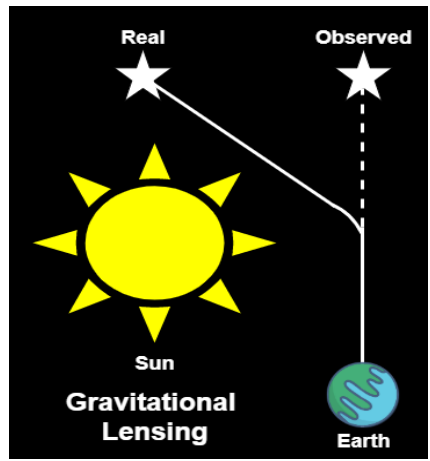}%&
\caption{The figure shows the phenomenon of Gravitational Lensing. Here the Sun is acting as a graviational lens and is curving the path of the light ray coming from a star behind it. Due to this phenomenon the light ray from that star is reacing the earth and the star is becoming visible.}\label{Gravitational Lensing}
\end{figure}

There are 3 types of gravitational lensing:-\\

1. Strong Lensing\\

2. Weak Lensing\\

3. Microlensing \\

In the strong lensing, the gravitational lens is generally very strong which causes it to create multiple images of the light source and sometimes it also results in formation of the Einstein ring. Einstein rings are formed when a very strong gravitational lens uniformly bends the light from the source. This generally happens when the light source, the gravitational lens and the observer lie on the same line. In weak Gravitational lensing, the lens is not so powerful to produce multiple images of the source. Microlensing is completely different from strong and weak lensing. Generally stars and exoplanets work as gravitational lenses in case of microlensing. Here the lensing effect between the very small exoplanet or stars and the massive light emitting celestial bodies which are very far from us causes the light source to appear very shiny for a period of time. Microlensing is used to detect the planets and stars which are very far from us and whose light becomes dim till they reach the Earth.  
}

\section*{\textbf{\Large{Gravitational Time Dilation}}} 

\large{Gravitational Time Dilation is the difference between the amount of time passed between two events measured by two different observers at varying distance from a gravitational mass. This is generally caused due to the effect of the gravitational field of a massive body in the 4-Dimensional spacetime. More the intensity of the gravitational field, more is the curving of spacetime. When the measuring clock is near the gravitational mass, the intensity of the gravitational field is more. Hence, the curvature in spacetime is also more. So light takes more time to traverse the curvature and the clock measures more amount of time. While in a free space, far from the gravitational mass, the measuring clock measures less amount of time as the light travels relatively faster now. This happens because the intensity of the gravitational field is very less and so it is not able to curve the spacetime.  In brief we can say that time runs slower when the effect of gravity is powerful.\\
\begin{figure}[ht]
\centering
\includegraphics[width=5.5cm, height=6cm]{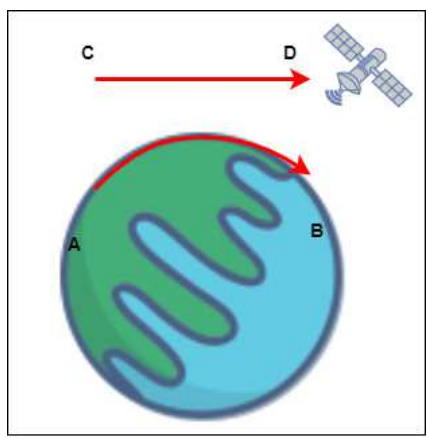}%&
\caption{The figure shows the path of light ray in space($CD$) and path of light ray when it passes near earth's surface($AB$). The light ray gets curved by the distorted space-time when it comes closer to the earth. }\label{Time Dilation}
\end{figure}

Mathematically this can be shown by:-\\
The length of AB $<$ Length of CD. \\

We know that light travels with a constant speed in all frames, that is $c$.$$speed = \frac{distance}{time}\hspace{2cm}[12]$$\\

As the speed is constant so to travel a larger distance, the light will take more time. So time taken in case of AB is more than in case of $CD$.  So, it states that light travels faster in free space and travels slower when it travels closer to a gravitational mass. This observation tells us that time is slower near a gravitational mass and time is faster in free space. This is the reason why the clock kept inside the international space stations runs faster than the clock kept on the surface of the earth.}

\section*{\textbf{\Large{Gravitational Waves}}} 

\large{The General Theory of Relativity had completely overturned the way in which Gravity functions. Einstein proposed that, when celestial objects having high masses accelerate, they disturb the space time in such a way that waves or ripples of these travel at the speed of light in all directions from the source of this colossal gravity, carrying the information about the source as well as the nature of gravity. These waves are invisible and can be imagined as the ripples generated in a pond when a stone is dropped in it. The detection and study of these waves are the key to taking the concepts of gravity up a level. Though the gravitational waves are weak in nature, the strongest of these waves are the ones which arise due to catastrophic events in the Universe, like the supernovae explosions, the collision of massive black holes, or possibly the Big Bang. Joseph Weber, in 1969, claimed to detect the gravitational waves using his self made gravitational wave detectors, now called as the Weber bars. However, due to inconsistency of data, these claims were reduced to insignificance. Around the same time, two scientists received the Nobel Prize for discovering Binary Pulsars, which helped in the detection of gravitational waves. Pulsars are the celestial objects rotating at a very high speed to emit electromagnetic waves. Binary Pulsars are pulsars with a binary partner. Over the next decade, pulsar timing observations were made, which proved that the withering of orbital periods corresponds to the loss of energy and angular momentum as explained by the general theory of relativity. This marked the first indirect evidence of gravitational waves. The direct evidence did not arrive until the year 2015. LIGO detected the gravitational waves using a laser interferometer for sensing the slightest signals of the weak gravitational waves arose due to the collision of black holes $1.3$ billion years away from the Earth. This discovery opened the controversial topic of the existence of gravitational waves and energy transfer to new ideas, additions and theories, completely changing the way we look at gravity, and even giving substantial evidence for the Big Bang.}

\section*{\textbf{\Large{Quantum Gravity}}} 

\large{There are four fundamental forces in nature. All four of them have been described in some or the other ways. While three of them have been explained using the Quantum field theory, Gravity is theorised using the general theory of relativity. For the unification of all the theories and removal of leftover anomalies, it is important to come up with a new arena of quantum gravity. It will ensure a surge towards the "\emph{Theory for everything}". Quantum Mechanics says that everything is composed of packets/small bundles known as quanta. These quanta can behave as particles as well as waves. For quantum gravity to flourish, the constituent particle needs to be discovered. This hypothesised gravitational quanta is known as the Graviton. Since quantum theories apply on strong fields and their effects are nearly impossible to measure in weak fields such as that of gravity, it seems to be a really tough path to unravel the connection between the quantum mechanics and relativistic theory of gravity. Many attempts have been made to clear the mist from these blended fields, the most successful ones being the M-theory and the loop theory. These theories have not yet completely been tested, but these theories have made remarkable efforts to integrate gravity with quantum mechanics Once quantum gravity is fully unravelled, the answers to all the questions would be known. And then would start a new era in physical sciences, with the formulation of the "\emph{Theory of Everything}".}

\section*{\textbf{\Large{Acknowledgement}}}
\large{We sincerely like to thank Dr. Tuhin Ghosh of School of Physical Sciences, NISER, for guiding us throughout the journey of writing these articles, helping us understand the formats and providing us with the important insights for writing a science article. Also, we thank him for reviewing our work, and helping us maintain the quality of the paper.}

%%References section

\end{document}